# On the Harmonic Oscillation of High-order Linear Time Invariant Systems

B. Baykant Alagoz

**Abstract:** Linear time invariant (LTI) systems are widely used for modeling system dynamics in science and engineering problems. Harmonic oscillation of LTI systems are widely used for modeling and analyses of periodic physical phenomenon. This study investigates sufficient conditions to obtain harmonic oscillation for high-order LTI systems. The paper presents a design procedure for controlling harmonic oscillation of single-input single-output high-order LTI systems. LTI system coefficients are calculated by the solution of linear equation set, which imposes a stable sinusoidal oscillation solution for the characteristic polynomials of LTI systems. An example design is demonstrated for fourth-order LTI systems and the control of harmonic oscillations are discussed by illustrating Hilbert transform and spectrogram of oscillation signals.

## Introduction

Linear Time Invariant (LTI) system modeling and analyses methods have been played an important role in development of science and technology for a century. LTI models have found widespread use in theoretical and numerical analyses of complex linear systems. Behaviors of dynamic systems are well characterized by LTI system models and consistency of LTI system analyses with real systems has been proven for numerous applied science and engineering problems. It is obvious that analyses on the base of LTI systems still play a central role in system analysis and system design, today. Deepened investigation on behaviors of LTI systems promises further contributions to modeling and comprehension of physical system behaviors.

LTI systems are expressed in the form of linear time-invariant differential equations. It is convenient to represent LTI differential equations as a set of first-order differential equations to perform state space analyses. Lablace transform providing transfer functions are mainly preferred because of simplification of the system analyses (Blackman, 1962; Perkins and Cruz, 1969; Harris and Miles,1980). The characteristic polynomial of the LTI systems provides a valuable tool for evaluating the character of LTI system (Gopal, 1993). Roots of characteristic polynomial are referred to as eigenvalues (or system poles) of the system. Complex conjugate roots ($\lambda_i = \sigma_i \pm j\omega_i$) lead to harmonic terms, $e^{jwt}$, in the solution of LTI models, and this results in harmonic components at system output (Gopal, 1993; Dorf and Bishop, 1990). The short-term oscillations, caused by the terms of $e^{\sigma_i t \pm jwt}$ for $\sigma_i < 0$, is transient and they asymptotically damps down in time. A continuous oscillating component at LTI system solutions appears when $\sigma_i = 0$ (Gopal, 1993; Dorf and Bishop, 1990). The current study is devoted for the design considerations of high-order LTI harmonic oscillators emerging in the cases of $\sigma_i = 0$ and $j\omega_i \neq 0$. Previously, stability boundary locus analyses were presented for the design of stable control systems (Tan 2005; Moghaddam and Abbasi, 2012). It is based on solving characteristic polynomial equation in s domain for $s = j\omega$ to figure out stability regions of controller coefficients. Indeed, stability boundary locus also indicates to the coefficients that results in oscillation of control systems. We consider it as an oscillation boundary and extend our investigation for the harmonic oscillation of LTI systems.

System oscillation is one of the most prominent problems of dynamic systems. Oscillation behavior of physical systems was extensively investigated due to potentials of modeling simplification and analyses consistency. Oscillation models are used for numerous application in various fields of applied science and engineering such as physics (Wells, 2012; Chiorescu et. al. 2004; Carloni et al. 2009; Yakovenko, 2006; Ibrahim and Tawfik, 2010), control science (Kasnakoglu, 2010), mechanics (Saha, 2012; Chen and Liu, 2009; Brigante and Sumbatyan, 2013), electrical system (Brigante and Sumbatyan, 2013; Radwan et al. 2008; Atay, 2002), biology (Wilkins et al. 2009; Jolma et al.,2010; Kaplan et al.2008). For instance, prevalence of harmonic oscillators in physics is obvious, describing small motions of an object attached to a string, molecules vibrating in crystals (Chiorescu et al. 2004). Oscillation conditions were stated for various oscillators (Wilkins et al. 2009; Radwan et al. 2008; Atay 2002). Investigation of conditions for harmonic oscillation (steady sinusoidal oscillations) of high-order LTI models may extend our understanding of complex physical systems, modeled by simple second-order oscillators, particularly in particle physics (Chiorescu 2004).

This study investigates sufficient conditions to obtain harmonically oscillating high-order LTI system model. Roots of the characteristic polynomials are mapped to amplitude- angle ($M - \theta$) plane by assuming roots of characteristic polynomial in the form of $\lambda = Me^{j\theta}$, and coefficients of the LTI system let to a desired harmonic oscillation were found by arbitrary solutions of linear equations. These arbitrary solutions impose sinusoidal oscillation solutions for characteristic polynomials. An example design was demonstrated for fourth-order LTI systems and, results were discussed for the control of harmonic oscillations.

**Methodology**

*Theoretical Background*

Differential equations of LTI control systems with constant coefficients $a_i \in R$ and a derivative order $n$ (Blackman, 1962; Perkins and Cruz, 1969; Dorf and Bishop, 1990) are expressed in a general form as follows,

$$\sum_{i=0}^{n} a_i y^{(i)} = u \qquad (1)$$

A LTI system is commonly expressed via state space representation (Gopal, 1993) as follows

$$x' = Ax + Bu \qquad (2)$$
$$y = Cx + Du, \qquad (3)$$

,where $x$ is $n \times 1$ state variable vector of the system and $y$ is system output vector. The tem $x'$ denotes the first derivatives of state variable vector. the $n \times 1$ vector $u$ represents input vector of the system. The matrix $A$ is $n \times n$ size state transition matrix of system. The matrix $B$, $C$ and $D$ models are model parameters of dynamic system. For single-input single-output systems, Lablace transform transfer function (Gopal, 1993; Abdelaziz, 2009; Özgören, 2009) was commonly expressed in s-domain domain as,

$$T(s) = \frac{Y(S)}{U(s)} = C(sI - A)^{-1}B + D \qquad (4)$$

Characteristic polynomial of this system (Gopal, 1993) were expressed as follows,

$$\Delta(\lambda) = \det(\lambda I - A) \qquad (5)$$

$\lambda_i \in C$  $i = 0,1,2,3,,n-1$ is complex eigenvalues of dynamic systems. In order to solve $\lambda_i$, let us express characteristic equation in polynomial form as follows,

$$\Delta(\lambda) = \sum_{i=0}^{n} \alpha_i \lambda^i = 0 \qquad (6)$$

The roots of characteristic polynomial yield $\lambda_i = \sigma_i \pm j\omega_i$ complex eigenvalues. For a zero input case ($u(t) = 0$), the time domain solution of LTI system contains sum of $e^{\lambda_i t}$ terms, $y(t) = a_0 e^{\lambda_0 t} + a_1 e^{\lambda_1 t} + .. + a_{n-1} e^{\lambda_{n-1} t}$ (Gopal,1993). Accordingly, the term $e^{\lambda_i t}$ can be decomposed product of two components as $e^{\sigma_i t} e^{j\omega_i t}$. Here, the terms $e^{j\omega_i t}$ yield harmonics at the output due to $e^{\pm j\omega_i t} = \cos(\omega_i t) \pm j \sin(\omega_i t)$ and results in sinusoidal components at the angular frequency of $\omega_i = 2\pi f_i$. The component $e^{\sigma_i t}$ specifies evolution of amplitude of $e^{j\omega_i t}$ in time. In the case of $\sigma_i < 0$, it is also commonly referred to as attenuation coefficient and correspondingly time constant of asymptotic attenuation was widely considered as $\tau_i = -1/\sigma_i$ for physical systems. These physical systems give response for a limited time period for instant disturbances. When $\sigma_i > 0$, the term of $e^{\sigma_i t}$ exponentially grows the amplitude of harmonics $e^{j\omega_i t}$ in time and results in instability of the system. If $\sigma_i = 0$, it stabilize amplitude of the harmonics with the frequency of $\omega_i$. It is well known that root locations in complex plane ($\sigma$, $j\omega$) tells us following remarks for high-order system behaviors (Blackman, 1962; Gopal, 1993; Dorf and Bishop, 1990),

(i) If all roots accommodates on the left half plane, so Re{$\lambda_i$} = $\sigma_i < 0$, $i = 0,1,2,3,,n-1$, the system will be asymptotic stability (Gopal, 1993). Because, all $e^{\sigma_i t}$ terms in solution go to zero, as the time goes infinity ($\lim_{t \to \infty} e^{\sigma_i t} \to 0$) (Gopal, 1993). If there is at least a root, of which $\omega_i \neq 0$, it yields a transient oscillation with time constant $\tau = -1/\sigma_i$.

(ii) If there is at least a root accommodating on the right half plane, so $\text{Re}\{\lambda_k\} = \sigma_k > 0$, $k \in [0, n-1]$, the system will be instable. Because, at least a term $e^{\sigma_k t}$ in solution goes infinitive as time goes infinity. If $\omega_k \neq 0$, it yields a instable oscillation with a growing amplitude.

(iii) If there are not a complex root at right half plane and at least a complex conjugate root accommodates on complex axis ($j\omega_i$); $\text{Re}\{\lambda_i\} = 0$ and for the rest of roots satisfies $\text{Re}\{\lambda_k\} = \sigma_k < 0$, $k \neq i$ and $k \in [0, n-1]$, this system oscillates. Because, the solution contains at least one harmonic term ($e^{jwt}$), which lead to sinusoidal oscillations in time domain solutions.

The system responses are represented with respect to root location of characteristic polynomials in Figure 1. Figure 2 illustrates simulation results obtained for sinusoidal oscillation stereotypes with respect to complex roots.

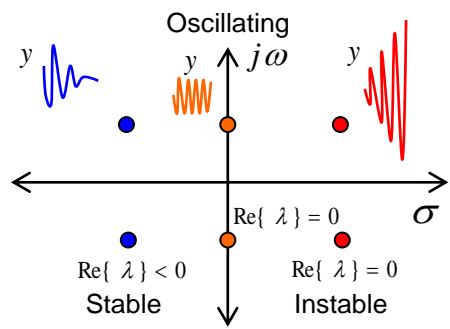

**Figure 1**. System responses with respect to root location of characteristic polynomial

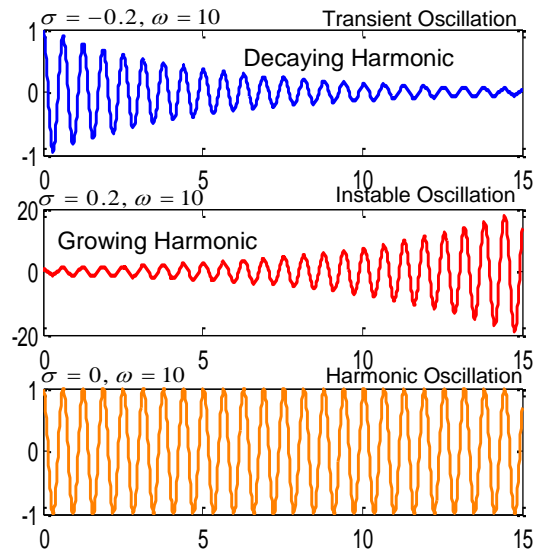

**Figure 2**. Oscillation of LTI system for various states of complex roots ($\lambda = \sigma \pm j\omega$)

## *Problem Definition*

This study aims to derive fundamental design considerations for harmonic oscillation of high-order LTI systems on the bases of characteristic root placement in complex plane. For this propose, lets express the roots of characteristic polynomial in the form of $\lambda_i = M_i\, e^{j\theta_i}$ (Gross and Braga 1961; Bayat and Ghartemani, 2008). Here, $M_i$ and $\theta_i$ are magnitude and angle of complex roots $\lambda_i$, respectively. Transformation from $(M,\theta)$ to $(\sigma, j\omega)$ is performed by using,

$$\begin{bmatrix}\sigma \\ \omega\end{bmatrix} = M\begin{bmatrix}\cos\theta \\ \sin\theta\end{bmatrix} \qquad (7)$$

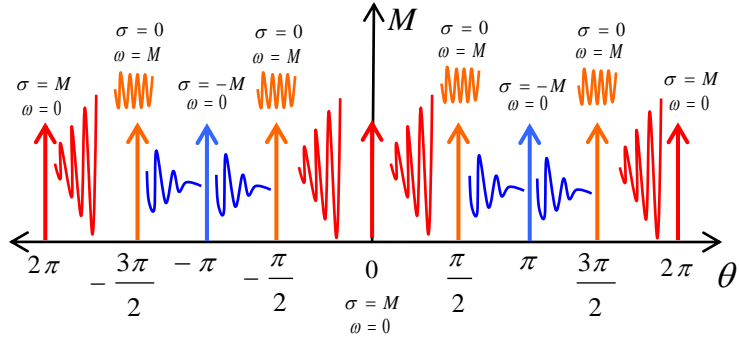

**Figure 3**. Root mapping in $M - \theta$ plane and resulting oscillations boundaries

Figure 3 illustrates the characteristics of systems in $M - \theta$ root mapping regions. According to figure, regions of $M > 0$ and $\frac{\pi}{2} + h\pi < \theta_i < \frac{3\pi}{2} + h\pi$ for $h \in T$ yield a asymptotically stable LTI systems due to coinciding to left half of complex root plane and regions of $M > 0$ and $-\frac{\pi}{2} + h\pi < \theta_i < \frac{\pi}{2} + h\pi$ for $h \in T$ yield instable LTI systems because of mapping right half of complex root plane. Boundaries between stability and instability mapping regions are referred to harmonic oscillating boundaries due to $\omega_i = M$ and $\sigma_i = 0$. Harmonic oscillation free response of LTI system is possible when all $\lambda_i$ is mapped over the boundaries of $\theta_i = \pi + h2\pi$ $i = 0,1,2,3,,n-1$ and $h \in T$ due to $\omega_i = 0$ and $\sigma_i = -M$.

Real physical systems in macro scales (mechanical systems) mainly exhibit attenuating harmonic oscillation ($\sigma_i < 0$) for a limited time interval depending on friction forces or thermodynamic energy loses. Time constant of this limited time oscillation (transient oscillation) is $\tau = -1/\sigma_i = -1/(M_i \cos(\theta_i))$ and the oscillating frequency is $\omega_i = M_i \sin(\theta_i)$. Steady oscillations can be mostly possible in micro-scale physical systems such as electro-physical and particle systems. However, instable harmonic oscillation ends with devastation or transformation of the physical systems and this type behaviors can be take place in very short intervals in the nature for instance short-lived particles.

## *Oscillation Boundaries of LTI systems*

*Property 1*: A LTI system yields steady oscillation if at least one root angle satisfies $\theta_i = \frac{\pi}{2} + h\pi$ $i = 0,1,2,3,,n-1$ and $h \in T$.

*Proof:* One can write the angle $\lambda_i$ as $\theta_i = \tan^{-1}(\omega_i / \sigma_i)$ and use it in $\theta_i = \frac{\pi}{2} + h\pi$. In this case, $\sigma_i$ can be written as,

$$\sigma_i = \omega_i / \tan(\frac{\pi}{2} + h\pi) \tag{8}$$

Since $\tan(\frac{\pi}{2} + h\pi) = \infty$ for $h \in T$, one obtains $\sigma_i = 0$ for $h \in T$. Considering remark 3, this LTI system oscillates due to $\lambda_i = \pm j\omega$.

*Property 2*: A n-order LTI system contains harmonic oscillation at the angular frequency $\omega_k$ for the first oscillation boundary ($h = 0$), when the characteristic polynomial coefficients $\alpha_0, \alpha_1, \alpha_2, ..., \alpha_n$ satisfy the following conditions,

$$\sum_{i=0}^{n} \alpha_i (\omega_k)^i \cos(\frac{\pi}{2}i) = 0 \quad \text{and} \quad \sum_{i=0}^{n} \alpha_i (\omega_k)^i \sin(\frac{\pi}{2}i) = 0$$

*Proof:*

If the characteristic equation of n-order LTI system given by equation (6) is solved for $\lambda_i = M_i e^{j\theta_i}$ mapping, we obtains,

$$\Delta(\lambda) = \sum_{i=0}^{n} \alpha_i (M_i e^{j\theta})^i = 0 \tag{9}$$

Considering Property 1 for the first oscillating boundary of system ($h = 0$), one can write the following equation,

$$\sum_{i=0}^{n} \alpha_i (M_i)^i e^{(j\frac{\pi}{2})i} = 0 \tag{10}$$

Since, $e^{j(\frac{\pi}{2})i} = \cos(\frac{\pi}{2}i) + j\sin(\frac{\pi}{2}i)$, equation (10) can be reorganized as follows,

$$\sum_{i=0}^{n} \alpha_i (M_i)^i \cos(\frac{\pi}{2}i) + \sum_{i=0}^{n} \alpha_i (M_i)^i \sin(\frac{\pi}{2}i) = 0 \tag{11}$$

For $\theta_i = \frac{\pi}{2}$, we can write $M_i = \omega_k$ and rearrange equation (11) as,

$$\sum_{i=0}^{n} \alpha_i (\omega_k)^i \cos(\frac{\pi}{2}i) + j\sum_{i=0}^{n} \alpha_i (\omega_k)^i \sin(\frac{\pi}{2}i) = 0 \tag{12}$$

A solution of equation (12) yields the sufficient conditions of harmonic oscillation at the angular frequency $\omega_k$ as,

$$\sum_{i=0}^{n} \alpha_i (\omega_k)^i \cos(\frac{\pi}{2}i) = 0 \quad \text{and} \quad \sum_{i=0}^{n} \alpha_i (\omega_k)^i \sin(\frac{\pi}{2}i) = 0 \tag{13}$$

Property 2 gives the sufficient conditions for determination of characteristic polynomial coefficients to make any n-order LTI system contain harmonic oscillation at the angular frequency $\omega_k$. One can write equation (13) in open form as, when the order n are even,

$$\alpha_0 - \alpha_2(\omega_k)^2 + \alpha_4(\omega_k)^4 - \alpha_6(\omega_k)^6 + ... \pm \alpha_{n-2}(\omega_k)^{n-2} \pm \alpha_n(\omega_k)^n = 0$$
$$\& \ \alpha_1 - \alpha_3(\omega_k)^3 + \alpha_5(\omega_k)^5 - \alpha_7(\omega_k)^7 + ... \pm \alpha_{n-3}(\omega_k)^{n-3} \pm \alpha_{n-1}(\omega_k)^{n-1} = 0 . \tag{14}$$

When the order n are odd,

$$\alpha_0 - \alpha_2(\omega_k)^2 + \alpha_4(\omega_k)^4 - \alpha_6(\omega_k)^6 + ... \pm \alpha_{n-2}(\omega_k)^{n-2} \pm \alpha_{n-1}(\omega_k)^{n-1} = 0$$
$$\& \ \alpha_1 - \alpha_3(\omega_k)^3 + \alpha_5(\omega_k)^5 - \alpha_7(\omega_k)^7 + ... \pm \alpha_{n-2}(\omega_k)^{n-2} \pm \alpha_n(\omega_k)^n = 0 . \tag{15}$$

These equations also tell us the following important remarks for oscillation condition of LTI systems:

(i) According to equation (13), due to arbitrary solutions, it is always possible to find out a characteristic polynomial coefficients set $\alpha_0, \alpha_1, \alpha_2.., \alpha_n$ to oscillate any LTI system at any $\omega_k$. (*Existence of solution*)

(ii) According to equation (13), the characteristic polynomial coefficients $\alpha_0, \alpha_1, \alpha_2.., \alpha_n$, which oscillates any LTI system at any $\omega_k$, is not unique. So, there is solution family of $\alpha_0, \alpha_1, \alpha_2.., \alpha_n$ satisfying equation (13). (*Abundance of solution*)

(iii) A special case of oscillating coefficients of characteristic polynomial appears when $\omega_k = 1$. A solution family can be obtained as $\alpha_0 = \alpha_1 = \alpha_2 = ... = \alpha_n$. (*Balanced solution families*) when the order n are even,

$$\alpha_0 - \alpha_2 + \alpha_4 - \alpha_6 + ... \pm \alpha_{n-2} \pm \alpha_n = 0 \text{ and } \alpha_1 - \alpha_3 + \alpha_5 - \alpha_7 + ... \pm \alpha_{n-3} \pm \alpha_{n-1} = 0 \quad (16)$$

when the order n are odd,

$$\alpha_0 - \alpha_2 + \alpha_4 - \alpha_6 + ... \pm \alpha_{n-3} \pm \alpha_{n-1} = 0 \text{ and } \alpha_1 - \alpha_3 + \alpha_5 - \alpha_7 + ... \pm \alpha_{n-2} \pm \alpha_n = 0 \quad (17)$$

(iv) Stability condition given by Equation (13) let to arbitrary selection of $n-1$ coefficient and the rest two coefficients are determined by the following equations,

$$\alpha_v = -\sum_{i=0 \& i \neq v}^{n} \alpha_i (\omega_k)^i \cos(\frac{\pi}{2}i) \text{ and } \alpha_g = -\sum_{i=0 \& i \neq g}^{n} \alpha_i (\omega_k)^i \sin(\frac{\pi}{2}i) \quad (18)$$

In order to avoid arbitrary locating of $n-2$ roots of characteristic polynomial and negative effects of this root on harmonic oscillation patterns, we prefer to locate them at $\theta_i = \pi$ with magnitudes $M_1 > M_2 > ... > M_{n-2}$, which mean to $\omega = 0$ and $\sigma_1 < \sigma_2 < ... < \sigma_{n-2} < 0$. Those roots yield transient asymptotic solutions ($e^{\sigma_i t}$) approximating to zero without any oscillation as time goes to infinity. In other words, we obtain $n$ number of linear equations for $n$ roots and choose only $\alpha_0$ arbitrary. Additional linear equation for non-oscillating transient roots can be written as,

$$\sum_{i=0}^{n} \alpha_i (\sigma_p)^i \cos(\pi i) = 0 \text{ for the each root non-oscillation root } (p = 1,2..n-2) \quad (19)$$

Design procedure for oscillating high-order LTI system ($n > 3$) with no transient oscillation can be summarized as follows,

Step 1: Set $\alpha_0 = 1$ or any real number.

Step 2: Write $n$ number of linear equations such that equation (13) is for oscillating roots and equation (19) is for each non-oscillation roots.

Step 3: Solve linear equations to obtain $\alpha_1, \alpha_2.., \alpha_n$ polynomials coefficients.

**Numerical Examples**

In this section, we present an example system for fourth-order LTI systems, expressed in the form of,

$$a_4 y^{(4)} + a_3 y^{(3)} + a_2 y^{(2)} + a_1 y' + a_0 y = u \quad (20)$$

In order to obtain harmonic oscillation component at $\omega_k = 2$ radian/sec according to Theorem 1, one writes equation (13) for $n = 4$ and then solves two equations given bellow,

$$\alpha_0 - 4\alpha_2 + 16\alpha_4 = 0 \text{ and } 2\alpha_1 - 8\alpha_3 = 0 \quad (21)$$

There are 5 design coefficients ($\alpha_0, \alpha_1, \alpha_2, \alpha_3$ and $\alpha_4$) and two linear equations, so one choose three coefficients arbitrary as $\alpha_0 = 1, \alpha_1 = 0.5$ and $\alpha_4 = 1$. Then, by solving equations (21), one obtains $\alpha_2 = 4.25$ and $\alpha_3 = 0.125$. Accordingly, the transfer function of oscillator can be obtained as,

$$T(s) = \frac{1}{s^4 + 0.125 s^3 + 4.25 s^2 + 0.5 s + 1} \quad (22)$$

State space model in controller canonical form can be written as

$$A = \begin{bmatrix} -0.1250 & -4.25 & -0.5 & -1.0 \\ 1.0 & 0 & 0 & 0 \\ 0 & 1.0 & 0 & 0 \\ 0 & 0 & 1.0 & 0 \end{bmatrix}, B = \begin{bmatrix} 1 \\ 0 \\ 0 \\ 0 \end{bmatrix}, C^T = \begin{bmatrix} 0 \\ 0 \\ 0 \\ 1 \end{bmatrix}, D=0 \quad (23)$$

A dirac function in the form of $u(t) = g\delta(t)$ was used for the input disturbance of the system. This imposes the initial condition of $y(0) = g$ and $y^{(i)}(0) = 0$, $i = 1,2,3..n$. The amplitude of dirac ($g$) can be used for adjusting the amplitude of harmonic oscillation.

Figure 4(a) shows roots ($\lambda_i, i = 0,1,2..4$) in complex plane. Complex conjugate roots $\lambda_{1,2} = \pm 2j$ yields harmonic oscillation at $\omega_k = 2$ radian/sec and the roots $\lambda_{3,4} = -0.0625 \pm 0.4961 j$ yield a transient decaying oscillation with a time constant $\tau = -1/(-0.0625) = 16$ sec at the angular frequency of $0.4961$ radian/sec. The transient oscillation almost attenuated in $5\tau \cong 80$ sec and the harmonic oscillation at $\omega_k = 2$ radian/sec continuous with an amplitude of 0.13 for $u(t) = \delta(t)$. In order to increase the amplitude of harmonic oscillation up to the amplitude of 0.65 as in Figure 4(c), $u(t) = 5\delta(t)$ was applied from the input. This figure demonstrate us harmonic oscillation amplitude for the high-order LTI system can be controlled via the input signal $u(t)$.

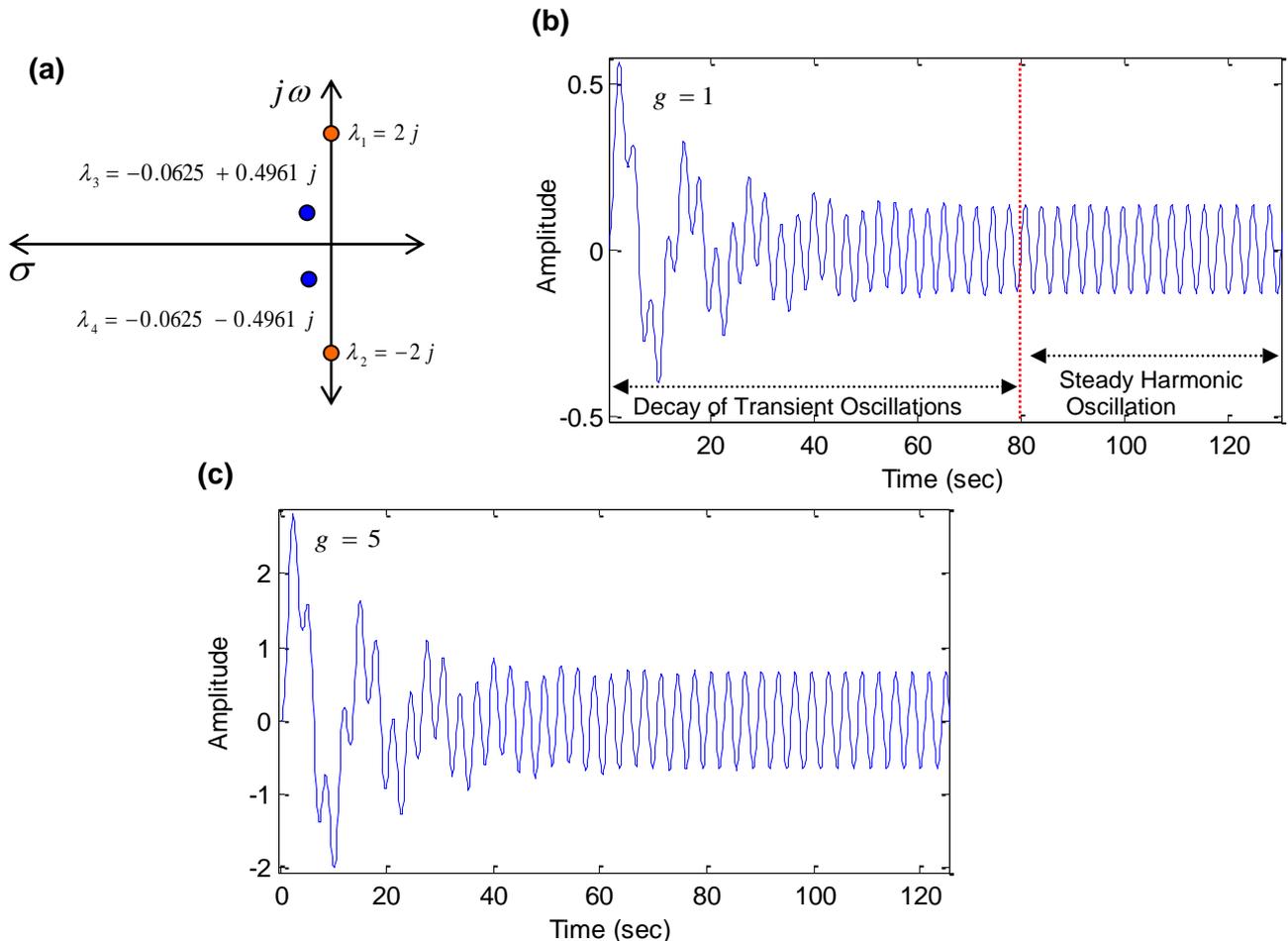

**Figure 4**. (a) Root locations in complex plane, (b) LTI system response for $u(t) = \delta(t)$, 8(c) LTI system response for $u(t) = 5\delta(t)$

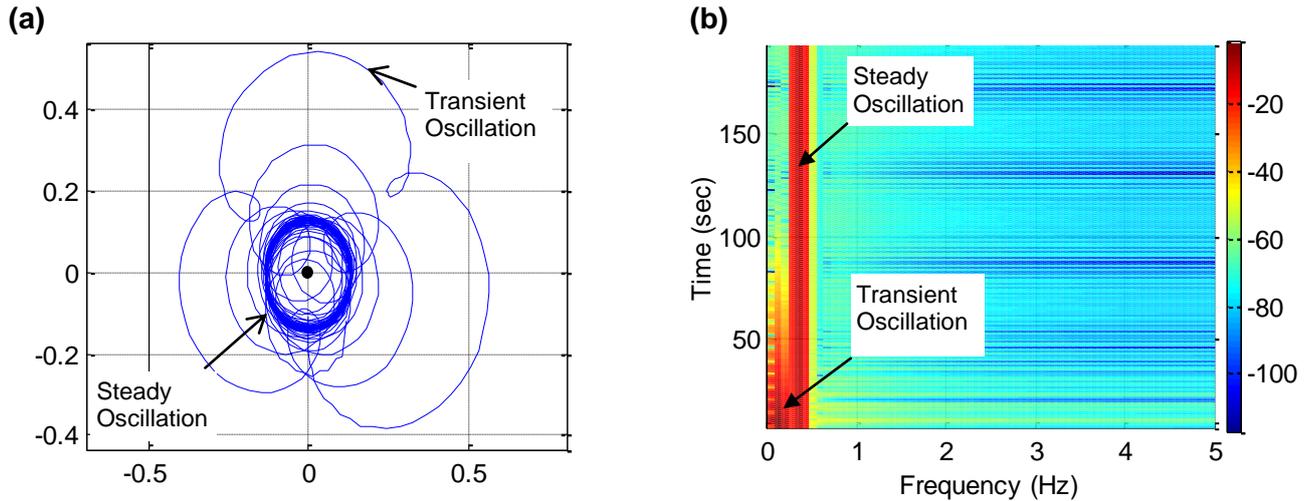

**Figure 5**. (a) Discrete-time analytic representation of the output signal. (b) Spectrogram of the output signal

Figure 5(a) shows discrete-time analytic signal of the system output calculated by Hilbert transform. Analytic signal represents signals in a complex plane ( $y(t) = |y(t)|e^{j\angle\phi(t)}$ ), where vertical and horizontal axis are complex part and real part of the complex signals. Irregular large circles were formed by transient oscillation in the figure. The circular orbit bands were formed by persistent harmonic oscillation obtained at $\omega_k = 2$ radian/sec. This effect can be clearly observed in Figure 5(b) demonstrating spectrogram of the system output obtained by moving window short-time Fourier transform. This figure shows transient damping oscillation at roughly 0.08 Hz (angular frequency of $0.4961$ radian/sec) in about 80 sec and a steady harmonic oscillation at 0.31 Hz (angular frequency of 2 radian/sec).

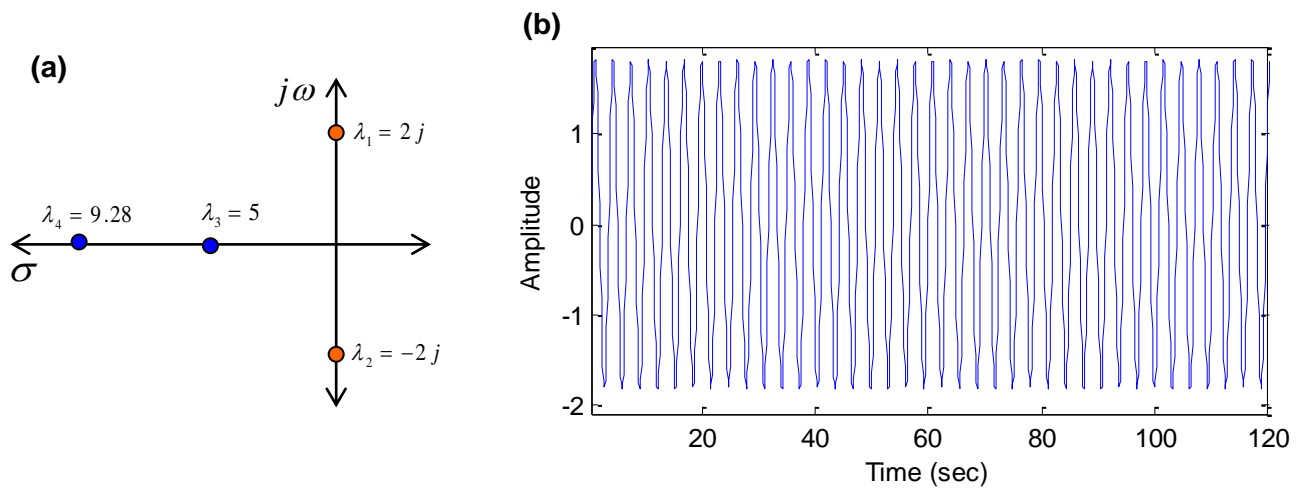

**Figure 6**. (a) Root locations in complex plane, (b) LTI system response for $u(t) = \delta(t)$,

In order to remove transient oscillations caused from the complex conjugates roots $\lambda_{3,4} = -0.0625 \pm 0.4961 j$, we applied the design procedure for oscillating high-order LTI system ( $n > 3$ ) with no transient oscillation given in previous section. For this proposes, we preferred decay of non-oscillating two transient components for $\sigma_3 = 5$ and $\sigma_4 = 9.2$ as in illustrated Figure 6(a). Two addition linear equations were written according to equation (19) as,

$$\alpha_0 - 5\alpha_1 + 25\alpha_2 - 125\alpha_3 + 625\alpha_4 = 0 \text{ and } \alpha_0 - 10\alpha_1 + 100\alpha_2 - 1000\alpha_3 + 10000\alpha_4 = 0 \quad (24)$$

For $\alpha_0 = 1$, the solution were obtained as $\alpha_1 = 0.3077$, $\alpha_2 = 0.2715$, $\alpha_3 = 0.0769$ and $\alpha_4 = 0.0054$ and the differential model of the system can be written as

$$0.0054 \, y^{(4)} + 0.0769 \, y^{(3)} + 0.2715 \, y^{(2)} + 0.3077 \, y' + y = u \quad (25)$$

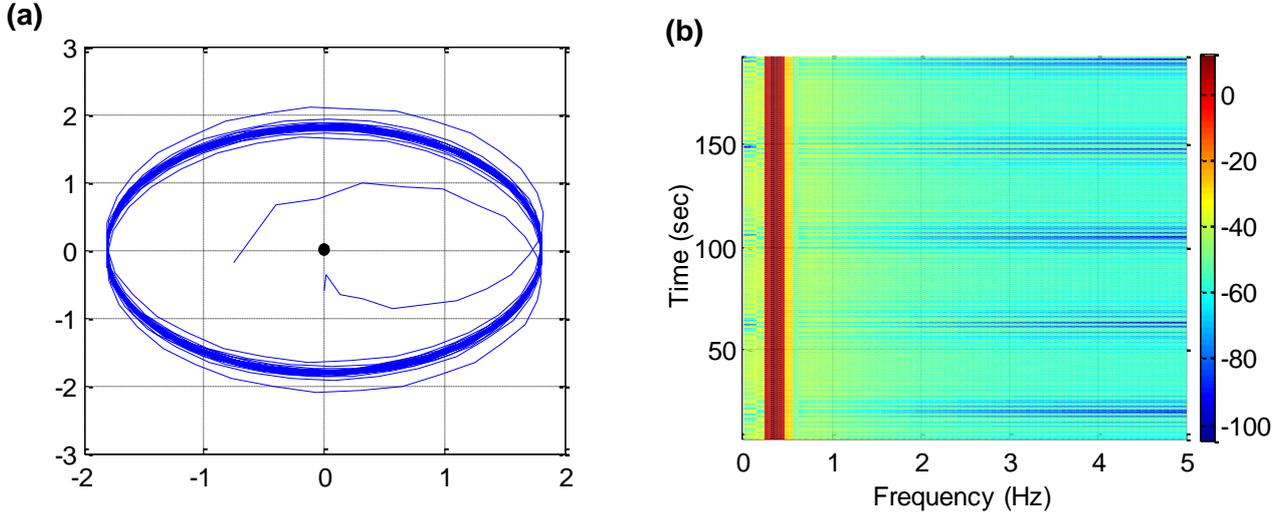

**Figure 7**. (a) Discrete-time analytic representation of the output signal. (b) Spectrogram of the output signal

Figure 6(b) shows the system output, where transient oscillations are cleared. To better evaluate harmonic oscillations, Figures 7(a) and 7(b) reveal discrete-time analytic signal and spectrogram of the system output. It is evidently seen from figures that transient oscillations were cleared from the system output. One of the noteworthy property of this oscillator is that one can add a bias level for harmonic oscillation by applying a step function ($u(t) = 1$, $t \geq 0$) to the input. Figure 8 shows step response and discrete-time analytic signal of the system. A bias with a magnitude of one is apparent in these figures. Figure 8(c) shows results of variable step input applied for a fourth-order LTI system designed for $\omega_k = 1$ radian/sec, $\sigma_3 = 5$ and $\sigma_4 = 9.8$. The differential model of the LTI oscillation control system was obtained as

$$0.0204 \, y^{(4)} + 0.3020 \, y^{(3)} + 1.0204 \, y^{(2)} + 0.3020 \, y' + y = u(t) \quad (26)$$

To better demonstrate amplitude and bias control of oscillations at $\omega_k = 1$ radian/sec via input signal $u(t)$, we applied subsequent delayed dirac input and a variable step input as control signal. The subsequent delayed dirac signal was formed as follows,

$$u(t) = 0.5\delta(t) + \delta(t - 100) + 1.2\delta(t - 150) \quad (27)$$

And, the following variable step input signal was used to disturb the oscillating system.

$$u(t) = \begin{cases} 0.5 & , t \leq 50 \\ 1 & , 50 < t \leq 100 \\ 2 & , 100 < t \leq 200 \end{cases} \quad (28)$$

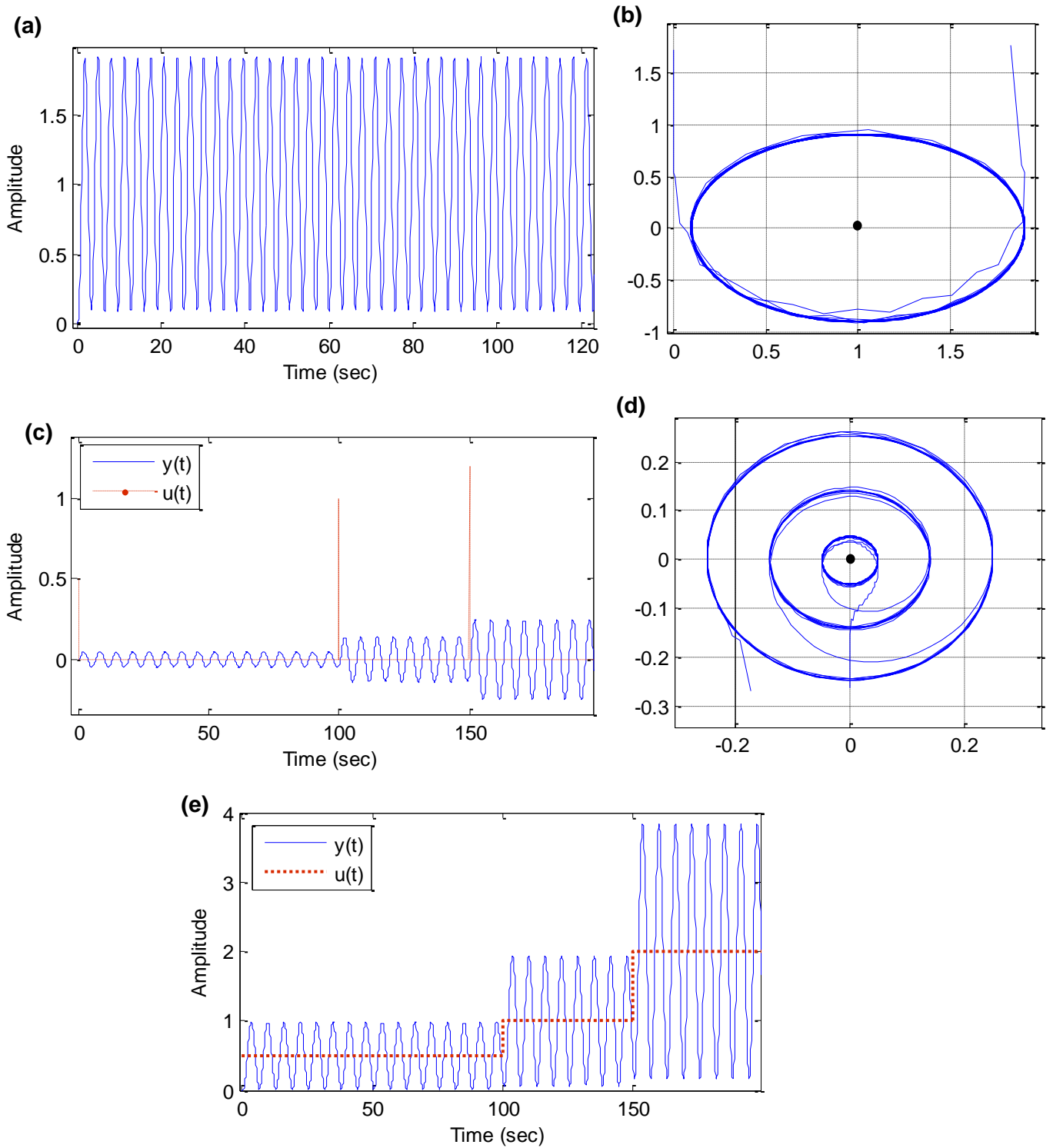

**Figure 8**. (a) LTI system response for $u(t)=1$ exhibiting a bias level of one, (b) Discrete-time analytic representation of the output signal with a bias level of one, (c) Harmonic oscillation controlled with a variable step input.

Figure 8(c) and (d) demonstrates control of oscillation amplitude by the subsequent delayed dirac signal generated by equation (27). Each pulse rises amplitude of harmonic oscillation with a zero bias. Figure 8(e) reveals shift of bias of oscillation by altering amplitude of the variable step function. In fact, this results implement an oscillation control function, which can be simply expressed by $y(t) = u(t)\sin(\omega_k t) + u(t)$.

**Conclusions**

This study presents harmonic oscillation control for high-order LTI systems. The paper discusses sufficient conditions of system coefficients to maintain harmonic oscillation and to clear transient harmonics from the system output. Moreover, some properties related to oscillating LTI system solutions such as existence of solution, abundance of solutions, balanced solution families are discussed.

Design and control examples were presented for the fourth-order LTI systems. In these examples, we illustrated removal of undesired transient harmonic oscillation from the system response and controlling the amplitude and bias levels of harmonic oscillations of the systems via input signal. Harmonic oscillation frequency can be adjusted by coefficients of the LTI system.

In physics, harmonic oscillations are widely used to explain mechanisms acting on the periodic events. Another noteworthy point should be emphasized that the steady state behavior of a high-order LTI system can mimic a second-order harmonic oscillator, when the complex conjugate roots resulting in transient oscillations moves away the complex axis. Steady-state response analysis of physical systems can hide effects of higher-order eigenvalues at the system output. However, the actual orders of harmonically oscillating systems can be estimated by evaluating transient response of systems.

Discussions and examples in the paper may be useful for system engineering and results can contribute to the field of vibration control and comprehension nature of oscillating physical systems as well.